\documentclass{sf2a-conf2016}
\usepackage{graphicx}
\usepackage{hyperref}
\usepackage[]{natbib}  
\usepackage{epstopdf}
\usepackage{caption}
\usepackage{subcaption}
\def\BibTeX{{\rm B\kern-.05em{\sc i\kern-.025em b}\kern-.08em
    T\kern-.1667em\lower.7ex\hbox{E}\kern-.125emX}}
\bibpunct{(}{)}{;}{a}{}{,}  

\begin{document}

\TitreGlobal{SF2A 2016}


\title{Modeling the reverberation of optical polarization in AGN}


\author{P. Andrea Rojas Lobos}\address{Observatoire Astronomique de Strasbourg, Universit\'e de Strasbourg, CNRS, UMR 7550, 11 rue de l'Universit\'e, 67000 Strasbourg, France}

\author{Ren\'e W. Goosmann\,$^1$}
\author{Fr\'ed\'eric Marin$^1$}



\setcounter{page}{237}


\maketitle


\begin{abstract}
According to the standard paradigm, the strong and compact luminosity of 
active galactic nuclei (AGN) is due to multi-temperature black body 
emission originating from an accretion disk formed around a supermassive 
black hole. This central engine is thought to be surrounded by a dusty 
region along the equatorial plane and by ionized winds along the poles. 
The innermost regions cannot yet be resolved neither in the optical nor in
the infrared and it is fair to say that we still lack a satisfactory 
understanding of the physical processes, geometry and composition of the 
central (sub-parsec) components of AGN. Like spectral or polarimetric 
observations, the reverberation data needs to be modeled in order to infer 
constraints on the AGN geometry (such as the inner radius or the half-opening 
angle of the dusty torus). In this research note, we present preliminary 
modeling results using a time-dependent Monte Carlo method to solve the 
radiative transfer in a simplified AGN set up. We investigate different
model configurations using both polarization and time lags and find 
a high dependency on the geometry to the time-lag response. For all models 
there is a clear distinction between edge-on or face-on viewing angles for 
fluxes and time lags, the later showing a higher wavelength-dependence 
than the former. Time lags, polarization and fluxes point toward a clear 
dichotomy between the different inclinations of AGN, a method that could help 
us to determine the true orientation of the nucleus in Seyfert galaxies.
\end{abstract}

\begin{keywords}
Galaxies: active, galaxies: nuclei, polarization, radiative transfer
\end{keywords}


\section{Introduction}

Active galactic nuclei (AGN) are the strongest steady sources in the Universe. 
While being extremely spatially compact, they produce enough bolometric 
luminosity to eventually outshine their host galaxy. It is an accepted 
paradigm that such strong and compact emission is due to accretion onto 
a supermassive black hole (SMBH, \citealt{salpeter1964,lyndenbell1969}). 
In thermal, radio-quiet AGN the SMBH is surrounded by an accretion disk 
producing the optical and UV continuum radiation (see \citealt{shields1978}, 
\citealt{ss1973} and \citealt{prp1973}). This emission is reprocessed by 
other structures surrounding the disk, such as the so-called broad line 
emission region (BLR), or a circumnuclear dusty medium often called 
the ``dusty torus'' \citep{Antonucci1993}. This dusty region extends to 
a spatial scale of a few parsecs for a $10^7 {\rm M}_\odot$ SMBH, but 
only in very few objects the circumnuclear dust is barely resolvable by 
near-IR interferometry techniques. The innermost parts cannot yet be 
resolved neither in the optical nor the infrared and it is fair to say 
that we still lack a satisfactory understanding of the physical processes, 
geometry and composition of the central (sub-parsec) components of AGN.

Both, the BLR and the dusty torus are somewhat confined to the equatorial 
plane that is defined by the accretion disk. The dusty torus is opaque to 
optical light \citep{Gaskell2009} and therefore obscures the BLR at all 
lines of sight intercepting it. Observers at such equatorial viewing angles 
do not see broad optical emission lines and therefore observe a so-called 
type-2 AGN. At polar viewing angles, the BLR is visible and the optical 
spectrum denotes a type-1 object. This is a fundamental axis of the 
so-called Unified Model of AGN that attempts to explain the observational 
diversity of active galaxies as an orientation effect. To verify this 
scenario the role of polarimetry was crucial. \citet{Antonucci1984} and 
\citet{Antonucci1985} observed polarized broad lines in highly inclined 
Seyfert galaxies and found a relation between low inclination (type-1) 
and high-inclination sources (type-2). They thus postulated that both AGN types 
share the same morphology but are seen at a different system inclination. 
Since then, a lot of effort has been put into understanding the complex 
polarization signal observed in the optical band. In particular, 
\citet{Smithetal2004} suggested that a flattened equatorial scattering 
region could explain the specific polarization observed in type-1 AGN: 
a low polarization degree $P$ (inferior to 1\% in most of the cases) 
associated with a polarization position angle parallel to the symmetry 
axis of the circumnuclear dust region. It was shown by our collaboration 
that also a wide half opening angle of the dusty torus produces the 
polarization characteristics of type-1 AGN \citet{Marinetal2012}.

Knowing that flattened scattering regions in type-1 AGN produce continuum 
polarization should make it possible to conduct polarization reverberation 
mapping as introduced by \citet{Gaskelletal2012}. The reverberation lag is due to 
a difference in light travel time between continuum radiation coming 
directly from the accretion disk and scattered radiation coming from 
structures farther away (e.g., from the torus, see Fig.~\ref{author1:fig1}, left). 
It is obtained by cross-correlating the optical continuum lightcurve with 
the polarized spectrum. The latter is obtained by multiplying the 
polarization fraction with the spectral intensity. The polarized time lag 
holds information on the average distance between the continuum source and 
the scattering regions. In this way, \citet{Gaskelletal2012} were able to 
infer the size of the inner scattering regions in the Seyfert-1 galaxy 
NGC~4151.

Like spectral or polarimetric observations, the reverberation data needs 
to be accurately modeled to infer constraints on the AGN geometry (such as 
the inner radius or the half-opening angle of the dusty torus). In this 
research note, we present preliminary modeling results using a time-dependent 
Monte Carlo method to solve the radiative transfer in a simplified AGN 
set up. We present results for polarization and time lags assuming different 
model configurations.

\section{Modeling the circumnuclear region}

\begin{figure}
  \centering
  \includegraphics[clip,width=0.44\linewidth]{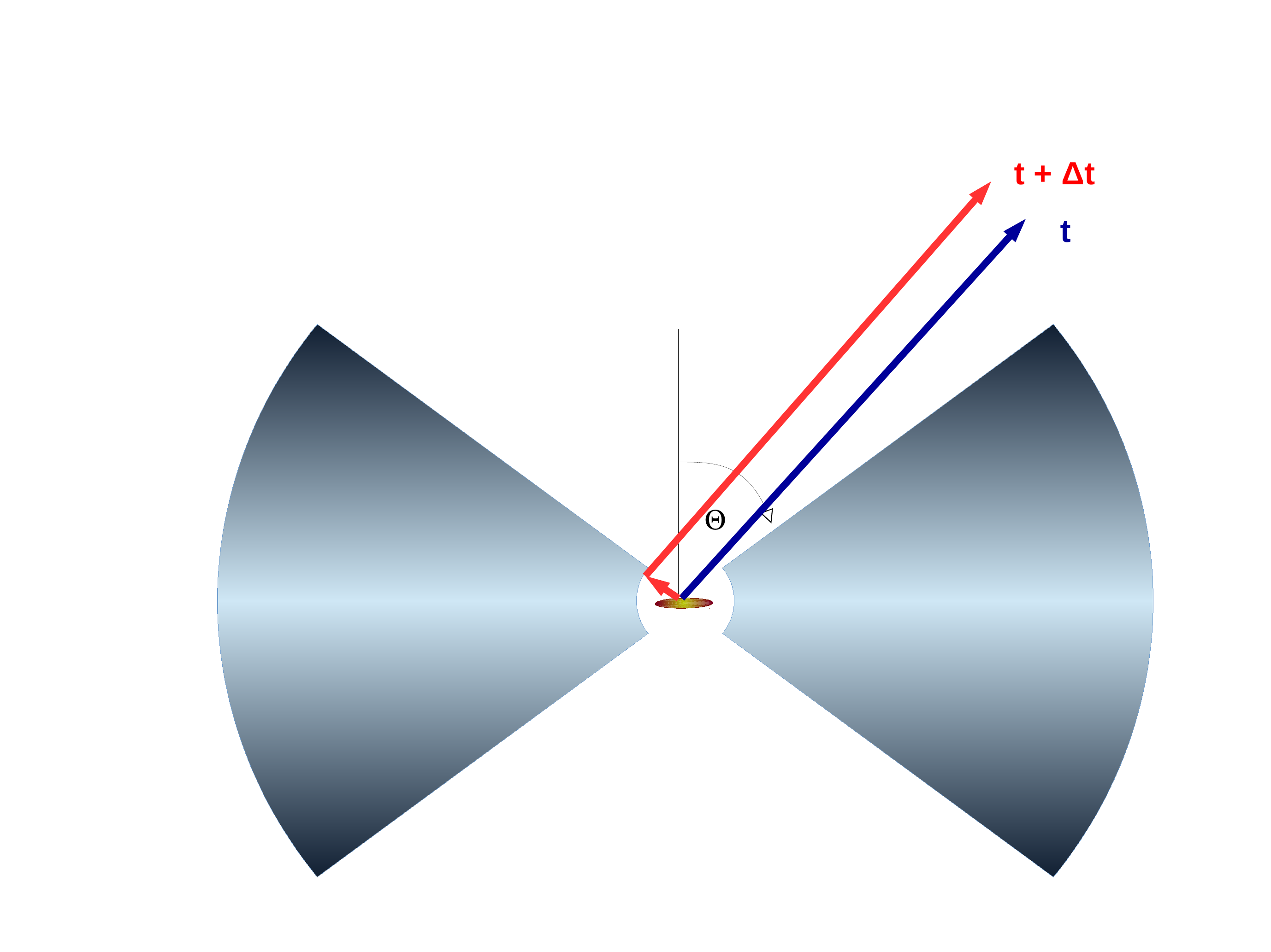}
  \hspace{1cm}
  \includegraphics[clip,width=0.44\linewidth]{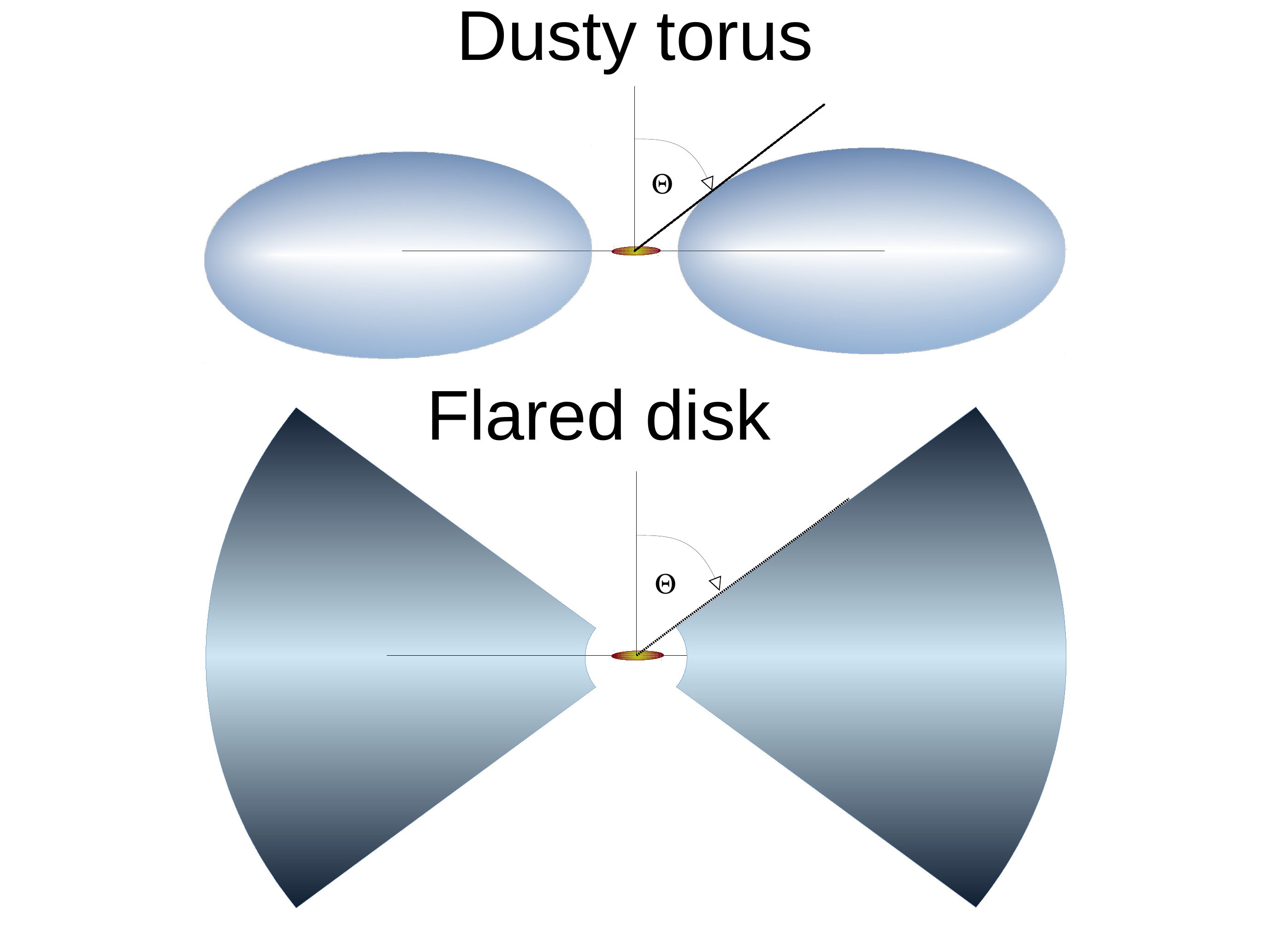}
  \caption{Left: Reverberation principle -- the blue line traces unpolarized photons coming directly from the source. The red line shows photons scattered inside the equatorial scattering region (polarized light). Right: model geometries -- the doughnut-shaped torus (top) and the extended flared-disk geometry (bottom).}
  \label{author1:fig1}
\end{figure}

The {\sc STOKES} code is a Monte-Carlo radiative transfer code written by \citet{GG2007}, \citet{Marinetal2012} and \citet{MGG2015} that computes the Stokes parameters of light, from which we can derive the polarization percentage, position angle, and total flux. In its latest public version, it also stores the time information in order to compute time-lags. The code allows the user to define different geometries and opacities for emission and scattering structures around a set of emitting regions. A free-to-download version of the code (v1.2 at this moment) can be found at: http://www.stokes-program.info/ . A recent review of the code performance is given in \citet{Marin2014}.

To investigate how polarized reverberation mapping can improve our knowledge of the morphology and composition of the dusty region at the center of AGN, we constructed a toy-model representative of the inner few parsecs of a Seyfert galaxy. At the center of the model, we implemented an irradiating continuum source using an isotropic point-like region emitting an unpolarized flux according to a power-law spectrum $F_{\rm*}~\propto~\nu^{\alpha}$ with $\alpha = 1$. Around it we defined a flattened dust distribution that could take two distinct forms: either a flared-disk or an elliptically-shaped torus. An illustration of the two geometries can be seen in Fig.~\ref{author1:fig1} (right). The filling of the equatorial region was either uniform (volume filling factor equal to unity) or non-uniform, i.e. clumpy (volume filling factor = 25). For all models, we considered the same torus dimensions: distance from the center: 0.0067~pc, external radius: 15.0067~pc, and inclination angle: 30$^\circ$ from the symmetry axis. The inner radius was set according to the time lag of 8 light-days found by \citet{Gaskelletal2012} for NGC~4151. The outer torus radius and the half-opening angle are based on near-infrared polarimetric observations by \citet{Ruiz2003}. The dusty region is optically thick, as is expected from observations with a total optical depth of $\sim$ 150 along the observer's line-of-sight. Both models (flared-disk and toroidal geometries) were tested with two dust prescriptions: ``AGN-dust'' presented in \citet{Gaskell2004} and ``MilkyWay-dust'' (``MW-dust'')  as prescribed by \citet{Mathis1977}. AGN-dust is composed of 85\% silicate and 15\% graphite; MW-dust is composed of 62.5\% silicate and 37.5\% graphite. The grain radius, $a$, varies from 0.005$\mu$m to 0.250$\mu$m following a distribution $n(a)~\propto~a^{s}$ with $s=-2.05$ for AGN and $s=-3.5$ for MW dust.

\section{Results}

\begin{figure}[t!]
 \centering
 \includegraphics[width=0.49\textwidth,clip]{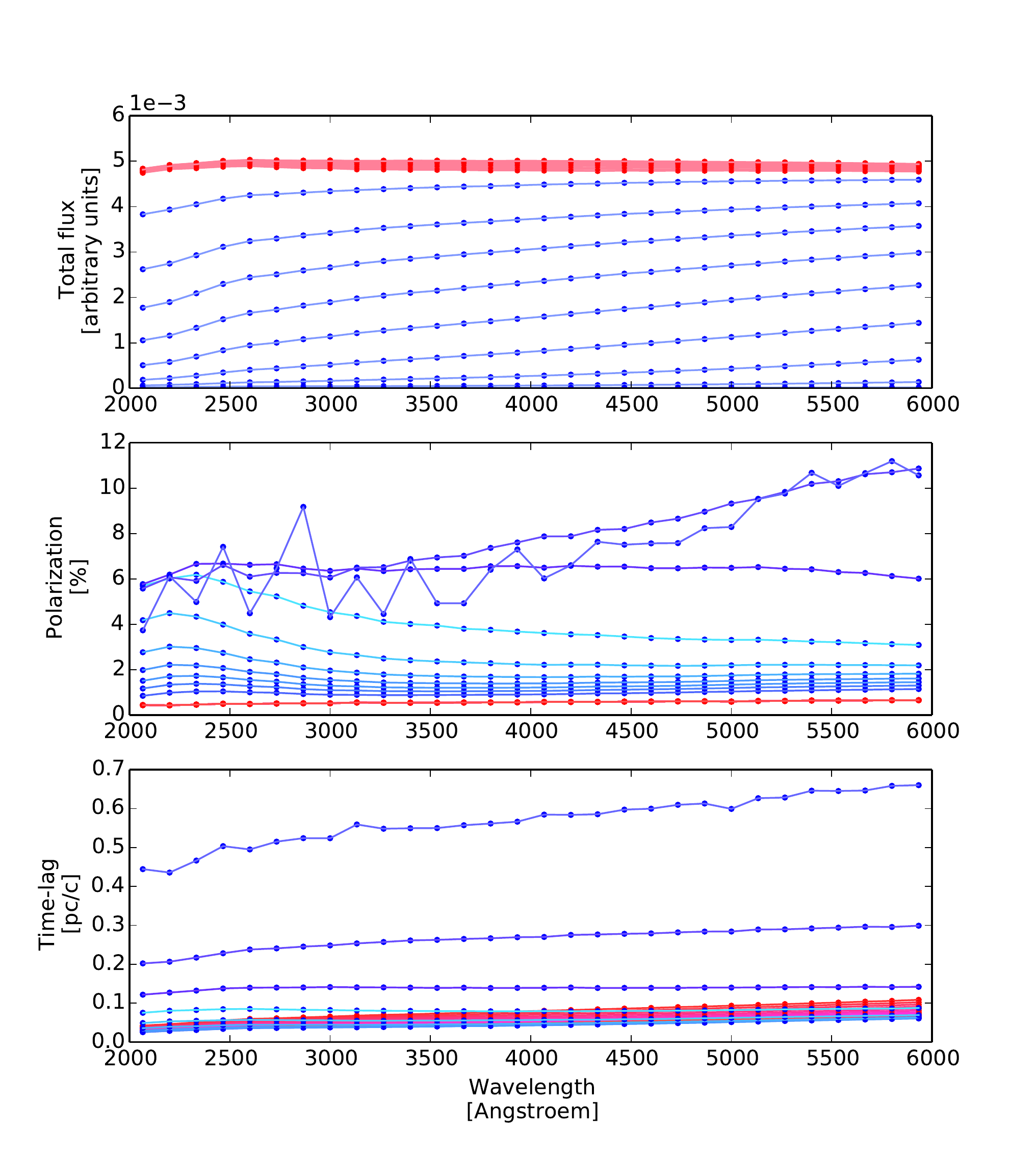}      
 \includegraphics[width=0.49\textwidth,clip]{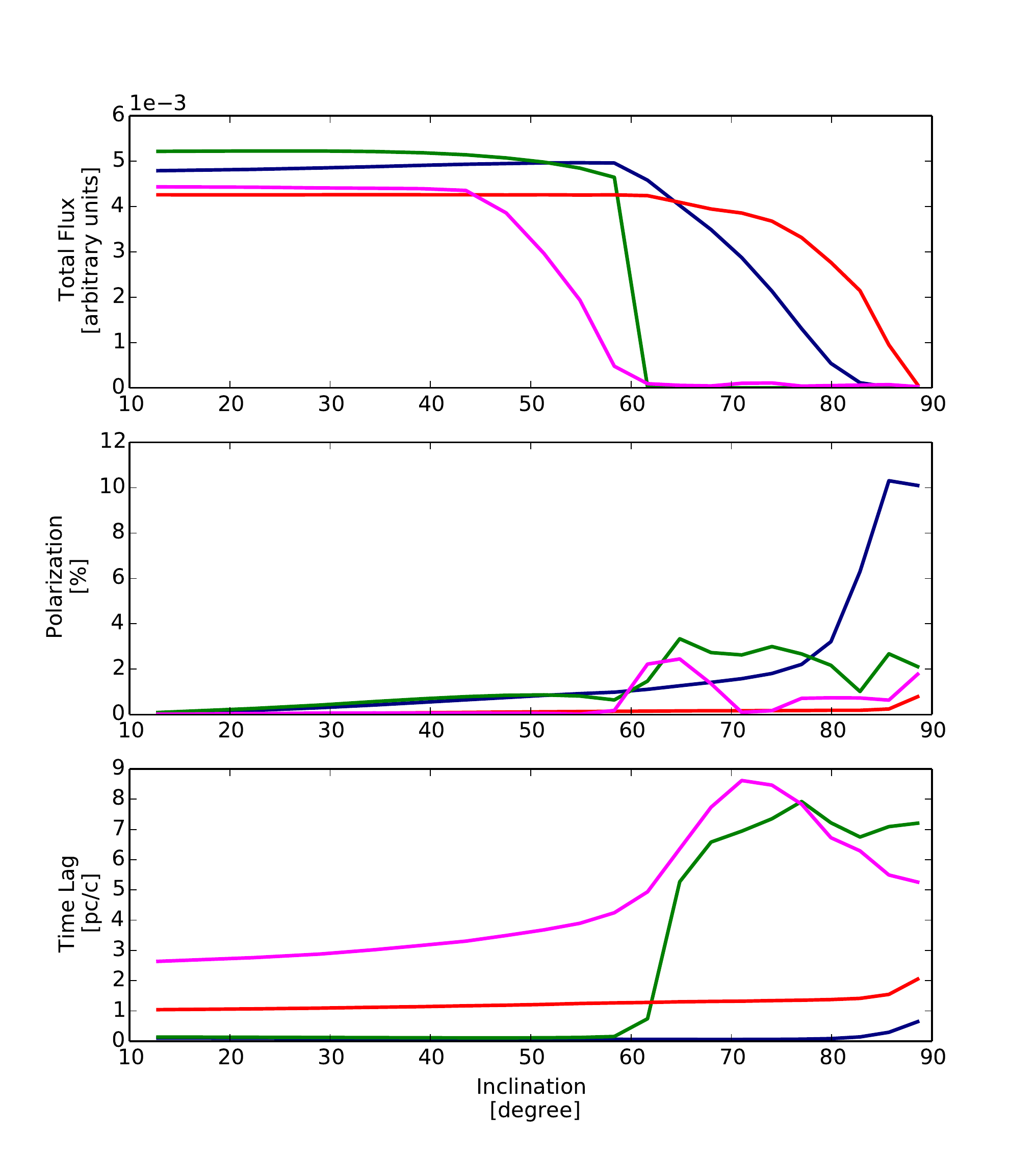}      
 \caption{Total flux (top), polarization percentage (middle) and time lag (bottom). The red dots show viewing angles lower than 60$^{\circ}$ (type-1) and blue dots show viewing angles above 60$^{\circ}$ (type-2). Left: Plots as a function of wavelength (left) for a uniform-density torus. Right: V-band data as a function of the system inclination. Blue line: uniform-dusty-torus, green line: uniform-flared-disk, red line: clumpy-dusty-torus, magenta line: clumpy-flared-disk.}
  \label{rojas:fig3}
\end{figure}

We first study the case of a uniform-density torus. In Fig.~\ref{rojas:fig3} (left), we plot the spectral flux, polarization and time lag as a function of wavelength. The spectral results can be compared to earlier work given in \citep{GG2007}. It turns out that both the strong direct emission seen at type-1 viewing angles and the significantly weaker scattered emission seen at type-2 orientations are mostly independent of wavelength. A slight depression is seen below 3000~\AA, which can be related to Mie scattering becoming more prominent versus Rayleigh scattering as well as to the presence of the dust-related 2175~\AA~feature. Apart from a similar variation in the near-UV and a slight overall rise towards longer wavelengths, the time lag is mostly insensitive to wavelength. This behavior is confirmed for the other modeling geometries studied in this work. Furthermore, it was discussed in \citep{GG2007} that the polarization properties induced by dust scattering in toroidal geometries do not vary much with wavelength. Next, we focus on the results as a function of the system inclination (Fig.~\ref{rojas:fig3} right). It turns out that the time lag of the polarized emission is not affected when modifying the dust prescription. Thus, we only present modeling results for the case of standard ``Milky Way'' dust. A discussion on the dust prescription and polarization effects is given in \citet{GG2007}.

The total flux and polarization as a function of inclination confirm what was shown in previous papers \citep{GG2007,Marinetal2012,MGG2015}. The flux must be much stronger at polar viewing angles (such as expected from type-1 AGN) than below the torus horizon (type-2) where only scattered radiation is seen. The scattered radiation is more strongly polarized but also accumulates a larger time lag. Notice that a time lag of zero corresponds to the direct emission from the source seen by the observer without any deviation. The fact that the time lag and the polarization fraction at type-1 inclinations are still larger than zero is due to the superimposed scattered component. At type-2 viewing angles the observed radiation must scatter inside the torus funnel before it escapes, increasing the time lag significantly.
 
The angular profile of the total flux, polarization and the time lag differ between the four geometries. We find a difference in the angular flux distribution between flared and toroidal shapes: the flared disk allows less radiation to be scatted towards type-2 viewing angles than the doughnut-shaped torus. A similar effect was found and discussed before when comparing the scattering properties between a large and a compact torus for the same half-opening angle \citep[see section~4.2 and figure~7 in][for more details]{GG2007}. For a given overall geometry, i.e. either flared or toroidal, the total flux and time lag are only in rough agreement with each other. Introducing clumpiness leads to a somewhat smoother and broader transition around the torus horizon and strongly lowers the polarization fraction and therefore the polarized flux.

\section{Summary and conclusion}
We modeled the polarization properties and time dependence of radiation scattered by circumnuclear dust in AGN. The inner radius of the dust region was fixed from polarization reverberation measurements in NGC~4151 \citep{Gaskelletal2012}. At type-1 viewing angles we find almost no variation in shape for all models, but we find a clear difference in the angular flux distribution depending on the geometry. Our results extend the work in \citet{GG2007} in terms of time lags.

There are additional differences in the polarization signature between a torus and a flared disk. While the torus models, either uniform or clumpy, do not show large differences in time-lag between type-1 or type-2 viewing angles, the case is different for flared geometries. The time lag is almost seven times larger at type-2 viewing angles than at polar inclinations. A flared disk structure leads to more important delays at edge-on inclinations, together with lower fluxes and polarization than for a toroidal dusty torus.

In conclusion, time-resolved polarimetry adds independent constrains to the unresolved structure of AGN and we aim to investigate in further detail these differences in a forthcoming paper.

\begin{acknowledgements}
This work was supported by CONICYT PFCHA/BecasCHILE Doctorado en el extranjero 72150573.
\end{acknowledgements}

\bibliographystyle{aa}  

%
\end{document}